\documentclass[preprint,showpacs,superscriptaddress,amsmath,amssymb]{revtex4}
\usepackage{dcolumn}
\begin{document}

\preprint{RPC}

\title{On Units for B(E2)'s}
\author{L. Zamick}
\affiliation{Department of Physics and Astronomy, Rutgers University, 
Piscataway, New Jersey 08854, USA}

\date{\today}

\begin{abstract}
In the table of Raman et. al. the units of B(E2) are given as e$^{2}$ 
barns$^{2}$ where e is the charge of the electron. Stelson and Grodzins 
however give B(E2) in units of length$^{4}$ (cm$^{4}$).
\end{abstract}

\maketitle


In several different sources the formula relating the transion probability 
per unit time for gamma decay from a 2$_{1}^{+}$ state to the $O^{+}_{1}$ 
ground state of even-even nuclei are written 
two different ways

\begin{equation} 
w_{1} = \frac{4\pi}{75} \frac{c}{\hbar c} (\frac{E}{\hbar c})^{5} B_{1}
(E2;2_{1} 
\rightarrow 0_{1}) 
\end{equation}

\begin{equation}
w_{2} = \frac{4\pi}{75} \frac{c}{\hbar c} (\frac{E}{\hbar c})^{5} e^{2} B_{2}
(E2;2_{1}\rightarrow 0)
\end{equation}

The equivilant of $w_{1}$ is given in the works of Blatt and Weisskopf [1] and 

J. D. Jackson [2].

The two formulae differ in where the squared charge of the electron i.e. 
$e^{2}$ is placed.  In $w_{1}$ the quantity $e^{2}$ is buried in $B_{1}(E2)$ 
and the 
dimension of $B_{1}$ is $e^{2}L^{4}$ where L is the unit of length that 
is used.  In $w_{2}$ the $e^{2}$ is lumped in with the other factors and 
the dimension of $B_{2} (E2)$ is just $L^{4}$.  We will use as our energy unit 

MeV, and our length unit fermi.  Note that $\hbar$ c = 197.32 MeV fm and 
$e^{2} = 1.44$ MeV fm, while $c = 3 \times 10^{23}$fm/s.
 
We then obtain:

\begin{equation}
w_{1} = 0.85 \times 10^{9} E^{5} B_{1} (E_{2}) 
\end{equation}

\begin{equation}
w_{2} = 1.23 \times 10^{9} E^{5} B_{2} (E_{2})
\end{equation}

Now in Raman's tables [3] one obtains the numerical value of the B(E2) from 
the meanlife by using the formula 

\begin{equation}
\tau (1+\alpha) = 40.82 \times 10^{13} E^{-5}/(B(E2, 0\rightarrow 2))
\end{equation}

Here the units are keV, barns and picoseconds where B is written (in our 
notation) as $B_{1}/e^{2}$.  But this is then simply $B_{2}$.  

Using the fact that $\tau = 1/w$, B(E2, 0$\rightarrow$ 2) = 5 B(E2,2 
$\rightarrow$ 0), 1 barn = 100 fm$^{2}$ and 1 keV = 10$^{-3}$ MeV we find 
that clearly the Raman formula is equivalent to $w_{2}$, i.e. that the factor 
of 
e$^{2}$ is not inside B(E2).  In the tables of Raman et.al. the 
dimensions of ``B(E2)'' are given as (e barn)$^{2}$.  In other words they have 

$B_{1}/e^{2}$ in formula (5) but quote $B_{1}$ in their tables.

The expression for $w_{2}$ is given in Heyde's book in 
Table 4 [5].  His units for B(E2), as he clearly states in eq. 4.23 and 4.27, 
are (fm)$^{4}$. On the 
other hand, in vol. 1 of Bohr and Mottelson [6] (see eq. 3C-18), and in 
Lawson's book [7] (see table 5-1 and eq. 5.12) they give the numerical 
values of $w_{2}$ (1.23) A very simple calculation 
shows that the coefficient of B in their transition 
rate is not 1.23 but 0.85 (times powers of 10).  They obviously have taken the 

$e^{2}$ out of the B(E2) (and multiplied this by 0.85). They perhaps 
confusingly use the same notation for B$_{1}$(E2) 
and for B(E2) in units of $e^{2}fm^{4}$.

We should point out that a compilation of B(E2)'s similar to that of Raman 
was made much earlier by Stelson and Grodzins [8].  Within powers of 10 the 
numerical values are the same as those of Raman.  Although not explicitly 
stated in the Stelson-Grodzins work, one can verify that they were using 
the $w_{2}$ relation to connect lifetimes to B(E2)'s.  And these authors state 

clearly in their work that the units of B(E2) are (Length)$^{4}$ -- they 
use cm$^{4}$.

\end{document}